# Transnational political violence in African borderlands

David G. Russell, University of Florida, david.russell@ufl.edu
Olivier J. Walther, University of Florida, owalther@ufl.edu

8/4/2026

**Abstract:** This paper examines the relationship between borderlands and political violence in Africa. Using spatiotemporal data on conflict events from 1997 to 2024 alongside the innovative OECD Spatial Conflict Dynamics indicator (SCDi), it suggests that borderlands see more conflict than non-borderlands. Political violence tends to decrease very rapidly with the distance to borders in Africa. The share of violent events recorded in borderlands varies over time and mirrors the cycles of violence that affect specific regions. It was very high in the early 2000s due to the wars in the Gulf of Guinea and Great Lakes region, declined for less than a decade, and increased substantially in the 2010s. Finally, the paper shows that when conflicts diffuse across borders, they tend to occupy a greater spatial extent, rather than relocate from one side of the border to the other.

## Violence in African borderlands

African conflicts are rarely confined to national borders. This tendency to spread transnationally has been evident since the end of the Cold War, first in the western Gulf of Guinea, with the conflicts in Liberia and Sierra Leone, then in the Great Lakes region, with the Congo Wars. After a short period of respite, the transnationalisation of violence took on a second wave in the early 2010s, with the outbreak of jihadist insurgencies around Lake Chad, in the Central Sahel, and in the Horn of Africa (Walther and Miles, 2018).

This spread of African conflicts across borders has been widely interpreted as a reaction to the weakness of states, which would allow criminal safe havens to develop on their peripheries, or as a manifestation of the principle of communicating vessels, whereby armed groups are pushed back from one territory to another by counterinsurgency initiatives.

Recent developments in African conflicts prompt us to revisit this state-centric explanation, which fails to take sufficient account of the strategies of non-state actors, such as transnational armed groups (Campana and Ducol, 2011). Indeed, recent years have shown that rebels and violent extremists compete with the state to control border regions and borderlanders (Cantens, 2022).

In some border regions, rebellious movements have established their own forms of governance, in an attempt to gain more autonomy from the central government, or develop a predatory economy based on the illegal extraction of natural resources (Scorgie, 2022). In other border regions, Jihadists groups have imposed an alternative political agenda that builds on a strict interpretation of religious law and challenges the very existence of modern nation-states (Thurston, 2020). States have reacted to this evolution by restricting cross-border mobility, conducting air strikes, and militarizing borders.

This paper examines the relationship between borderlands and political violence at the continental level. Building on disaggregated conflict data, the paper first examines the presumed link between borders and conflict by analysing whether political violence tends to cluster in borderlands or whether it is rather more spatially dispersed across the continent. Previous studies conducted in North and West Africa suggest that violence tends to decline substantially with the distance from borders (OECD, 2022), but it remains unclear whether a similar relationship is observed elsewhere in Africa.

The paper then mobilizes temporal data to examine whether African borderlands have become more violent than other regions over the last 27 years. The existing literature suggests that African conflicts follow certain regularities in the way violence is temporally distributed (Walther et al., 2025) but the contribution of border conflicts to these cycles or waves of violence remains understudied. Finally, the paper examines how armed conflicts diffuse, asking whether conflicts tend to relocate or expand across international borders in Africa. While the process of diffusion is well documented in the literature, it has rarely been applied to understand the spatial dynamics of conflicts at the continental level.

The paper suggests that borderlands see more conflict than non-borderlands and that the occurrence of violent events decreases strongly with the distance from borders in Africa. The strength of this relationship hides considerable temporal and spatial variations, however. Some borderlands are more conflict-prone than others, and the level of violence in these borderlands tends to rise and fall in sync with broader waves or phases of conflict across Africa.. Furthermore, conflict in borderlands is more clustered and more intense than conflicts elsewhere. When these conflicts diffuse across borders, they tend to expand geographically, rather than relocate from one side of the border to the other (Figure 1).

Figure 1. Diffusion by expansion versus diffusion by relocation

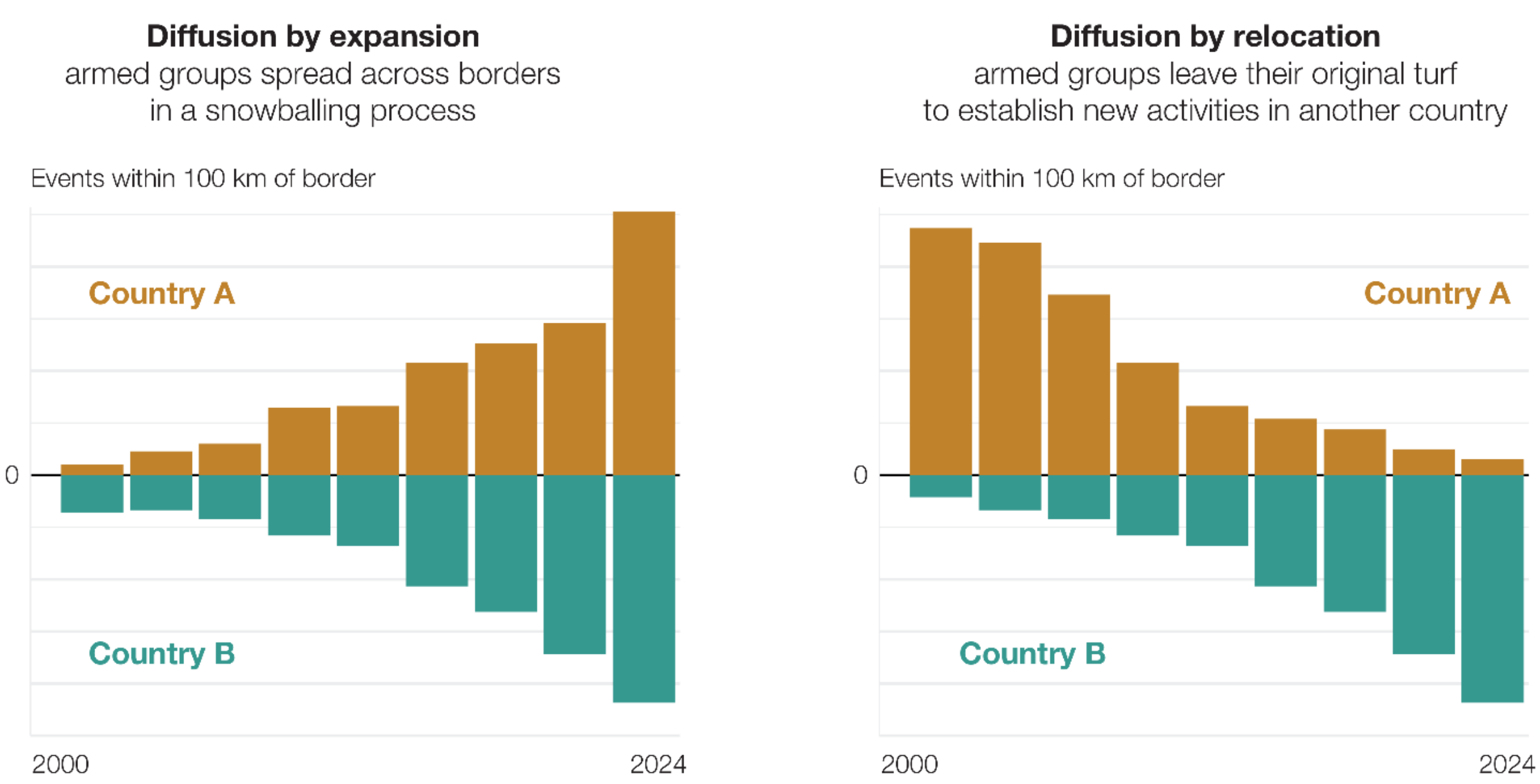

## Literature review

*Borderlands and violence*

African borderlands have experienced an increase in violence since the end of the Cold War (Salehyan, 2011). This "transnational turn" has been interpreted  in binary terms: states are either too weak to prevent violent organisations from operating within their territory, or strong enough that they force these groups to relocate through counterterrorism strategies (Bøås and Jennings, 2007).

The first approach argues that states who have lost their monopoly on the legitimate use of violence tend to attract violent organisations that are either expelled from other countries or in search of a more secure haven (Howard, 2010). As weaker states become safer havens for transnational terrorism, this contributes towards spreading violence to neighbouring countries. The idea that failed states allow violent organisations to spread in ungoverned areas and create sanctuaries has grown in popularity after the collapse of Somalia and the Congo Wars (Ghani and Lockhart, 2009).

The second approach to African border disorders (Walther and Miles, 2018) attempts to explain the relocation of violent extremist organisations by the impact of counterterrorism strategies. Accordingly, borderlands can provide favourable grounds to violent organisations defeated in their home country by strong states, who have no other choice but to relocate across borders. In other words, violent extremists cross borders out of necessity rather than out of opportunity. D'Amato (2018: 156), for example, notes that states' weaknesses should be seen as "an enabling condition that cannot be assumed to be the causal explicative factor".

International expansion is a carefully orchestrated decision since it implies considerable physical, social, and strategic costs for violent extremist organizations (Skillicorn et al., 2021). Crossing borders is also time consuming and potentially dangerous when insurgents move to an unfamiliar environment where they can no longer rely on civilian support. These costs associated with border crossings explain why even the most "transnational" groups tend to operate largely within the limits of one country (Dowd, 2018; OECD, 2022).

*Hotspots of borderlands violence*

State and non-state actors tend to cross borders for various reasons that impact borderlands differently across the continent, creating major hotspots of violence, such as the Central Sahel, around Lake Chad, in the Great Lakes region, and in the Horn of Africa (Walther et al., 2025). Government forces have been known to cross into neighbouring countries to restore order, by cutting insurgents' communication lines and establishing military bases. Intervening in a neighbouring country can also help create a buffer zone and prevent incursions of violent extremists, as when Kenya and Ethiopia attacked Al-Shabaab's strongholds in Somalia in 2011. State forces may also go transnational to coordinate a regional offensive against a common enemy. The Multinational Joint Task Force (MNJTF) established around Lake Chad is a recent example of this strategy (Emmanuel, 2018).

Non-state actors tend to spread violence across borders when they relocate after being defeated by government forces. In Central Africa, the Lord's Resistance Army, historically based in northern Uganda, moved to the Democratic Republic of Congo (DRC) and the Central African Republic (CAR) after a series of military offensives by Ugandan forces (Schomerus, 2021). Non-state actors also use borderlands to recruit, train and plan attacks against more distant targets, as during the civil wars of Sierra Leone and Liberia (Reno, 2011). Finally, rebels and extremists can also move to borderlands to exploit the local grievances that decades of under-investment and political marginalization have created in the peripheries of the state. The expansion of Jihadist groups affiliated with Al Qaeda and the Islamic State along the borders of Burkina Faso and Niger in the Sahel is the most obvious example of this strategy (Bøås et al., 2020).

*Spatial diffusion of conflict*

The spatial clustering of political violence does not necessarily mean that conflict itself is contagious (Buhaug and Gleditsch, 2008). Instead, conflict hotspots could also be explained by the fact that other variables that lead to conflict, such as ethnic tensions or income inequality, are also clustered in specific regions. These critiques have led to two different perspectives among recent inquiries into the diffusion of conflict.

In the first camp of research, scholars have described the patterns of diffusion that conflicts might take across space. While *relocation* characterises a phenomenon that occurs at one place and then moves to another place, *expansion* refers to a phenomenon that takes place over a greater and greater spatial extent (Lohman and Flint, 2010). Braithwaite and Jeong (2017) point to distinguishing between two types of spatial diffusion of conflict across borders as a promising avenue for future research. The first type is learning, or emulation, where actors near to a conflict emulate the violent behaviours they see perpetrated nearby. The second type is spillover, where conflict between the same actors expands to occur across a greater spatial extent. When civil wars diffuse, they tend to expand in spatial extent rather than to relocate from one area to another (Schutte and Weidmann, 2011).

The second camp of research comprises studies that examine the conditions under which conflicts spread spatially, with a focus on states as the primary spatial units of analysis. In this framework, war can spread to involve more states based on those states' geographic proximity to the states involved in the war (opportunity) and their diplomatic relationships with them (willingness) (Radil et al., 2013). Gleditsch et al. (2008) investigate four ways that civil wars can lead to international conflict involving the state experiencing the civil war, finding that the most common mechanisms are outside intervention into the civil war, externalization of the civil war to retaliate for a neighbour's harbouring of rebel groups, and spillover due to effects like refugee migration and the damage of infrastructure in borderlands.

These mechanisms have been examined by Buhaug and Gleditsch (2008), who note that while civil wars tend to cluster in time and space in general, conflict contagion is more likely when ethnic groups in one state are involved in a separatist conflict in another state. Kathman (2010) argues instead that civil wars spread geographically because neighbouring states are motivated to intervene in nearby civil wars due to a threat of more widespread instability. Metternich et al. (2017) extend this finding to show that a combination of geographic proximity and presence of excluded ethnic groups in both the sender and receiver states increases the likelihood of conflict diffusion the most. Political violence can cluster in time and in space because of how nearby conflicts change local popular acceptance of the legitimacy of the use of political violence, as Linke et al. (2015) show in their study of 16 African countries.

A key question in determining the mechanisms by which conflicts diffuse is investigating the specific role of borders. Black (2013: 752) contends that, with a limited definition of conflict diffusion, "in the vast majority of cases violent civil conflicts do not spread across borders". In their global study of conflict contagion, Metternich et al. (2017) note that some borders do stop conflict contagion, particularly when the receiver state enjoys a high level of state capacity. Skillicorn et al. (2021) model this phenomenon at a granular spatial scale, considering borders alongside simple geodesic distance as an impedance to conflict diffusion from the point of view of a commander of an armed group.

**Box 1. Methodology**

This paper uses data from Armed Conflict Location & Event Data Project (ACLED, 2024), which records instances of political violence and mobilization throughout the world as disaggregated event data. Each entry provides data on the date and georeferenced location of the event, the type of event, the actors involved, fatalities (if any), sources used to create the data, and other relevant information (Raleigh et al., 2023).

The analysis focuses on armed conflict in African states that have a land border with at least one other state. Only battles, explosions or remote violence, and violence against civilians that occurred from January 1, 1997, through December 31, 2024, are included. ACLED data for the resulting 48 states for this time period includes 213,126 unique events.

To simplify the analysis and ensure replicability across the continent, borderlands are defined as areas within 100 km of an international land border. Using R's sf package, we calculated the distance between each ACLED event and the nearest border, recording also which border was nearest. Counts of conflict events both within 100 km of borders and outside of that buffer area were aggregated by distance to nearest border, by year, by region, and by nearest border.

To examine the spatial patterns of political violence inside and outside of borderlands, we turned to the Spatial Conflict Dynamics indicator (Walther et al., 2023), which uses a 50km-by-50km grid overlaid onto event data. For each grid cell and for each year within a given time period, the SCDi calculates the spatial clustering (measured with an Average Nearest Neighbour Index or ANN) and spatial intensity of events of political violence. Clustering scores are then classified into clustered (ANNI < 1) or dispersed (ANNI >= 1), and intensity values are classified into high or low relative to a generational mean. This gives a set of four possible SCDi values for each year depending on whether violence is more or less intense and more or less clustered.

**Borderlands are more violent than other spaces**

Borders are often the sites of intense conflict as actors take advantage of state vulnerabilities or flee from state strength. At the continental level, conflict data collected since 1997 indicate that borderlands are indeed much more violent than other regions. Graphing the occurrence of events of political violence by distance to the nearest international border shows a clear distance-decay effect (Figure 2) where more conflict tends to occur in borderlands than would be expected if events were randomly distributed in space.

This trend is not universal, however, and certain conflicts are concentrated far away from borderlands. This happens when conflicts are mainly limited to one state, when that state's borderlands are relatively sparsely populated, and when conflict has primarily centred around urban areas in the core of the state, as has been the case in Somalia, Libya, and Sudan. The large spike in events around 300 km from a border visible on Figure 2 is almost exclusively driven by the series of conflicts that have centred around Mogadishu, Somalia.

Figure 2. Violent events by distance to nearest border in Africa, 1997-2024

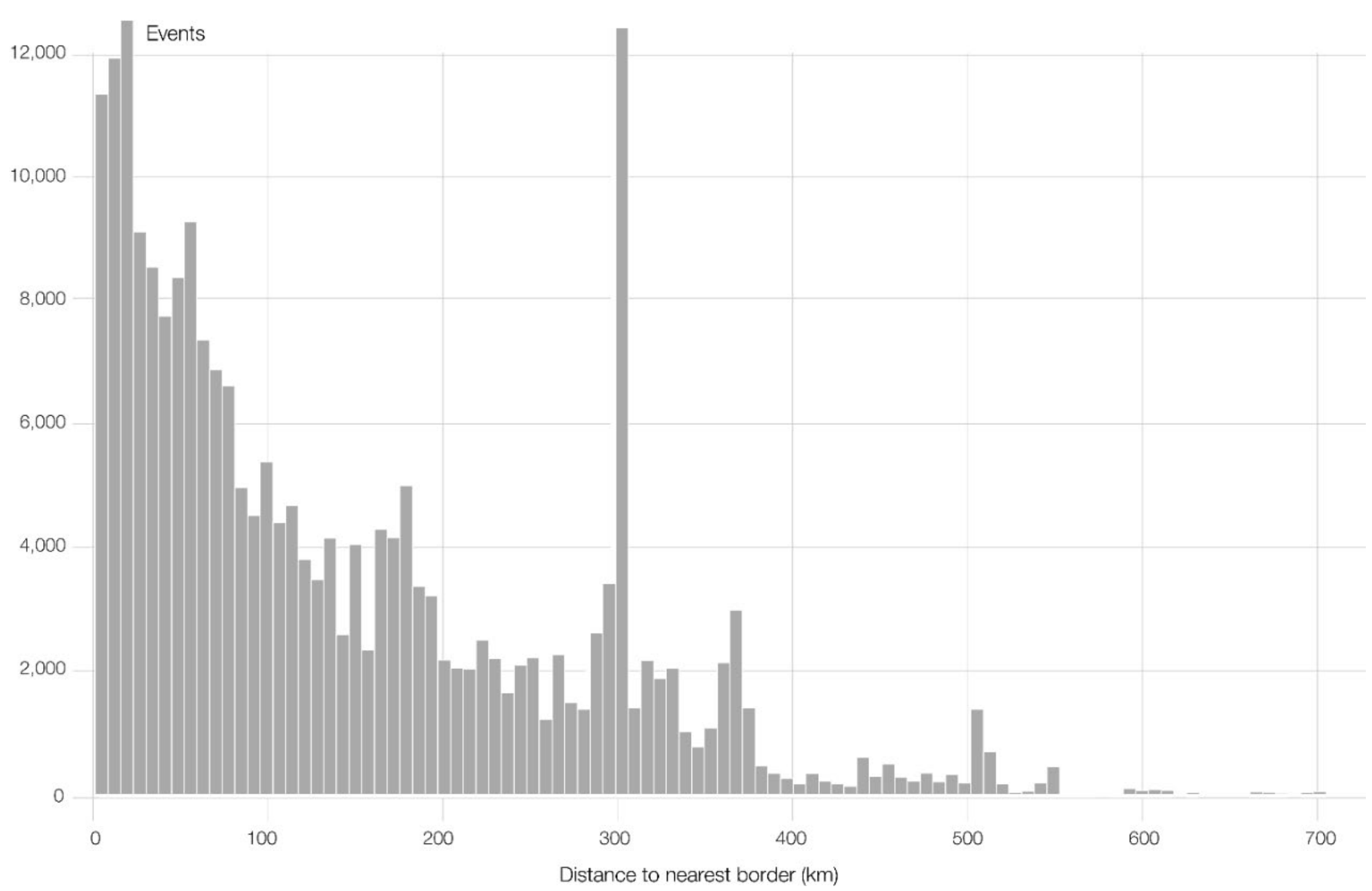

Source: ACLED (2024), calculations by the authors.

There is significant regional and temporal variation in the clustering of conflict near to borders in Africa. On average, around 50% of conflict events from 1997-2024 occurred within 100 km of an international border: the median distance to a border was 99 km, while the mean distance from a border was 140 km. These results are in line with previous studies conducted in North and West Africa, which suggested that 46% of all violent events occurred within 100 km of a border (OECD, 2022; Radil et al., 2023). Everywhere on the continent, the first few kilometres near a border are far more violent than the rest of the border region or the interior of the country.

The relationship between violence and borders has varied significantly over the last 27 years (Figure 3). These variations reflect several waves of violence, as conflicts emerge, expand, and disappear in several regions of the continent. The percentage of violent events within 100 km of borders was particularly high until in the early 2000s in most regions, due to several conflicts that very explicitly built on border regions, as in Liberia, Sierra Leone, Guinea, Côte d'Ivoire, and DRC. This period of high concentration of violence was followed by a few years of relative decline in most regions except Middle Africa, which stands out for its consistently high concentration of conflict events near to borders since 2000. The Second Congo War and its violent aftermath have resulted in a landscape of political violence within the DRC that is highly focused on its eastern borders with Uganda, Rwanda, and Burundi.

Figure 3. Violent events within 100 km of borders in Africa, by region, 1997-2024

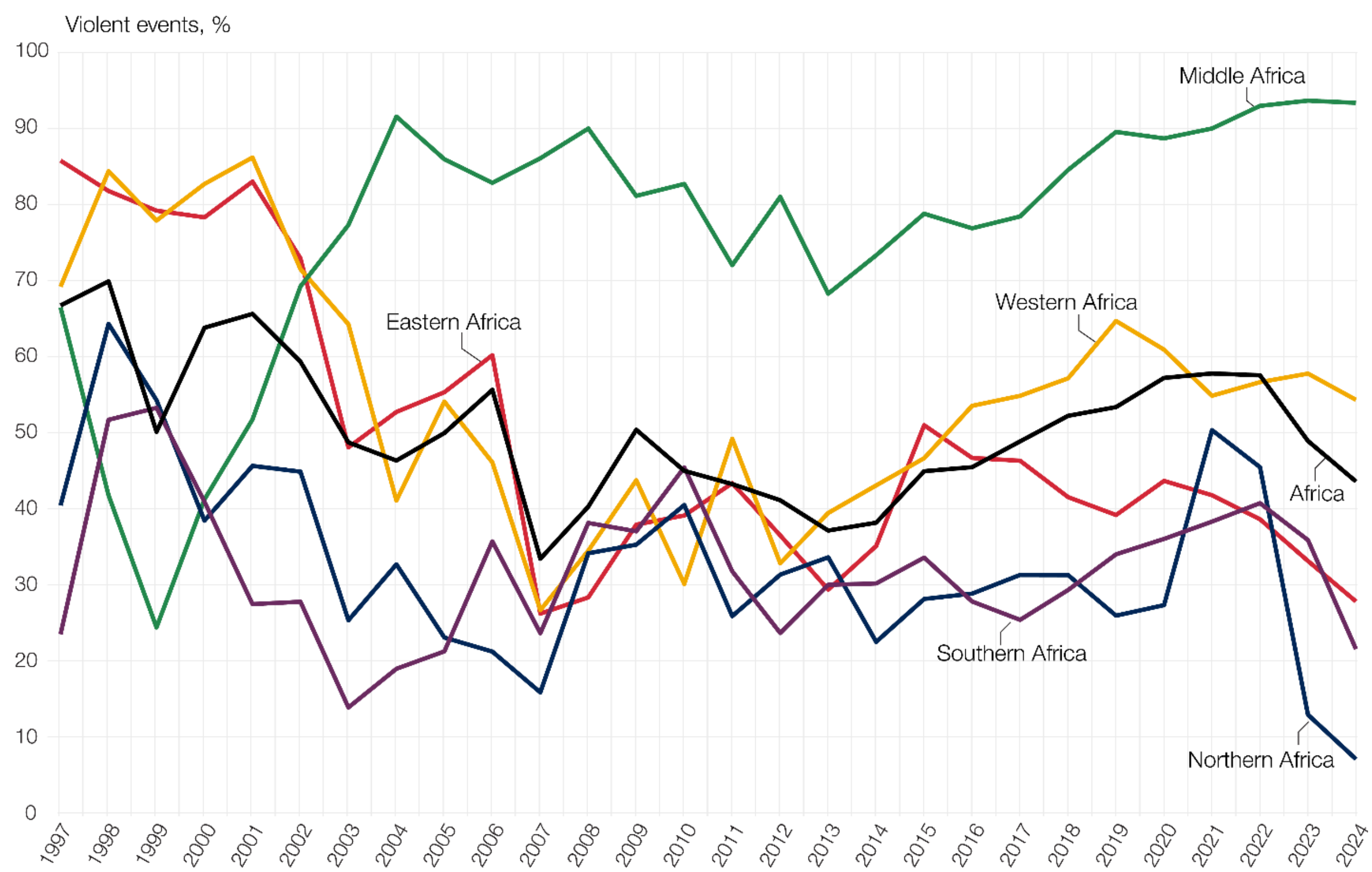

Source: ACLED (2024), calculations by the authors. See Figure 3 for how countries are sorted into regions.

The broad continental trend of increasing closeness to borders from 2012-2022 is driven primarily by conflict dynamics in Western Africa. The initial increase of borderlands conflict in Western Africa coincides with the onset of the 2012 civil war in Mali. Conflict events in Nigeria, Mali, and Burkina Faso continue to occur near to borders, though the interior of the Nigerian state has seen more violence since 2019. The significant downward trend in the percentage of conflict events near to borders in Africa in recent years is driven by conflict dynamics in both Northern and Eastern Africa. In Northern Africa, this trend is particularly strongly driven by the intensifying conflict in Sudan, and to a lesser extent, by the lingering conflict in Libya, where a permanent ceasefire was signed in 2020. In both of these conflicts, actors have primarily focused their efforts on a handful of urban areas in the cores of the contested states, rather than in border regions. In the Eastern African context, nearly half of these conflict events have occurred in Somalia, where conflict patterns are centred around Mogadishu.

**Violence tends to cluster in some borderlands**

African conflicts are very unevenly distributed across the continent and tend to affect only a few border regions. To highlight which border regions are the most affected by violence, we first use the SCDi, which combines the *intensity* of violence with its spatial *concentration*. Combining these two variables is crucial to distinguish different stages of conflict and understand how they are perceived locally by civilian populations: two regions may experience the same number of violent events, while resulting in vastly different locational patterns, some more concentrated or diffuse than others.

The results suggest that violence in borderlands is more intense and more clustered on a local scale than in non-border areas in Africa (Figure 4). The indicator shows that most cells in borderlands experience clustered and intense (type 1) violence, strongly associated with intensifying and persistent conflict (Walther et al., 2023). More dispersed or less intense types of conflicts can be found in the peripheries of border regions, which suggests that borderlands correspond to the core of many conflict regions in Africa. More than half of conflict cells in borderlands experience clustered and high-intensity violence, more than in the other regions (Table 1).

Table 1. Proportion of conflict types in borderlands and other regions in Africa, by type, 1997-2024

| SCDi conflict types | Borderlands | Other |
|---|---|---|
| Type 1. Clustered/high intensity | 56% | 52% |
| Type 2. Dispersed/high intensity | 2% | 2% |
| Type 3. Clustered/low intensity | 33% | 37% |
| Type 4. Dispersed/low intensity | 9% | 9% |

Source: ACLED (2024), calculations by the authors.

Figure 4. Conflict types in Africa, 2024

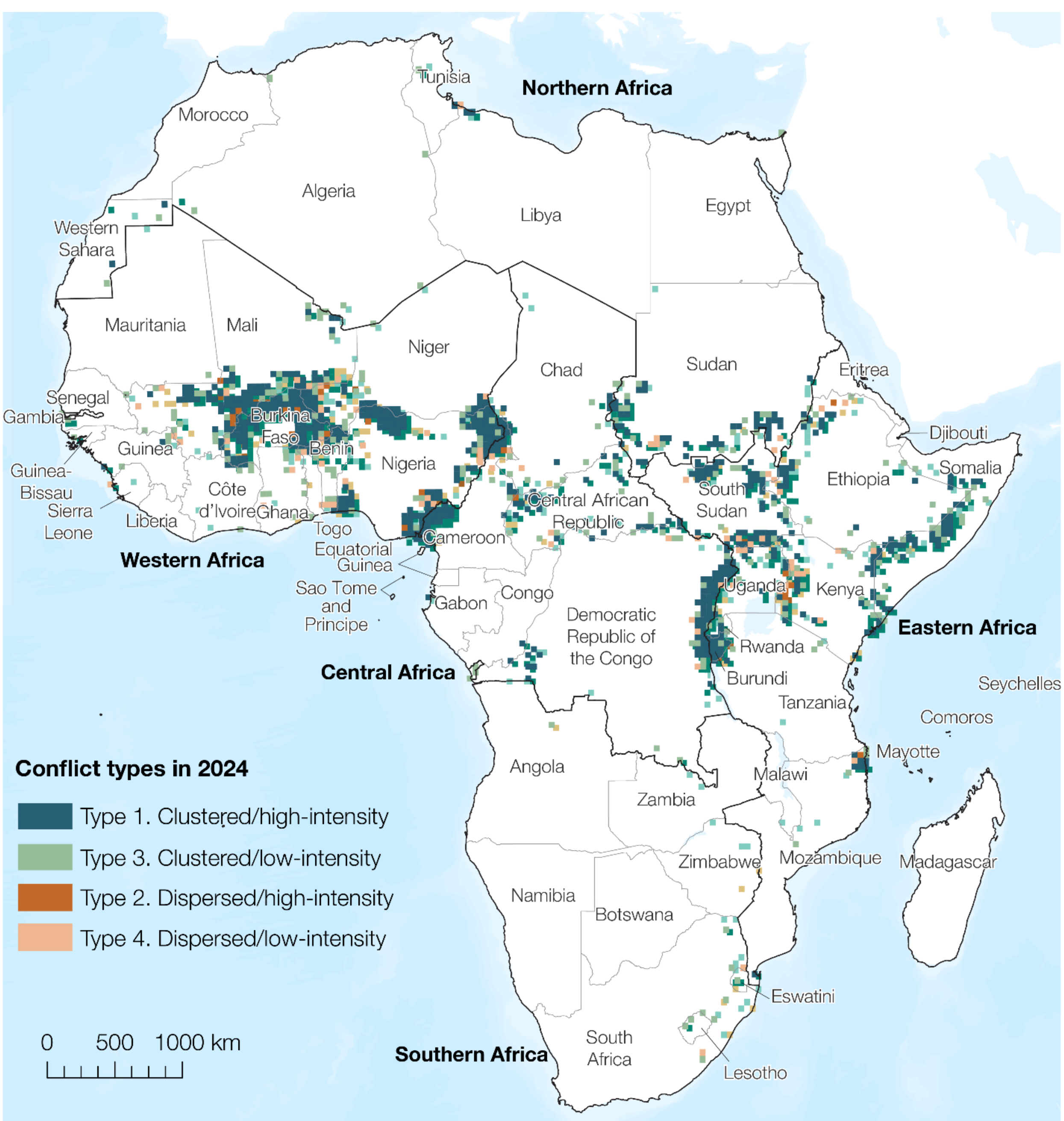

Source: ACLED (2024), calculations and cartography by the authors. The indicator is mapped for grid cells within 100 km of an international border.

In addition to being strongly clustered in certain border regions, political violence also affects specific segments of African borders more than others. To visualise this unique geography, we identified the nearest border segment to each violent event and measured the distance between the events and nearest border segment. These data were then used to mapwhich African border segments were most affected by political violence (Figure 5). Using violent events rather than fatalities ensures a higher degree of accuracy in this analysis, since the number of people killed is often subject to debate, while the occurrence of an event is rarely disputed. However, this choice may lead to underestimates of some (interstate) conflicts, such as the one between Ethiopia and Eritrea, which has caused massive casualties without necessarily resulting in many events, due to the heavily mechanized forms of warfare adopted by the belligerents.

Figure 5. African border segments by number of violent events within 100km, 1997-2024

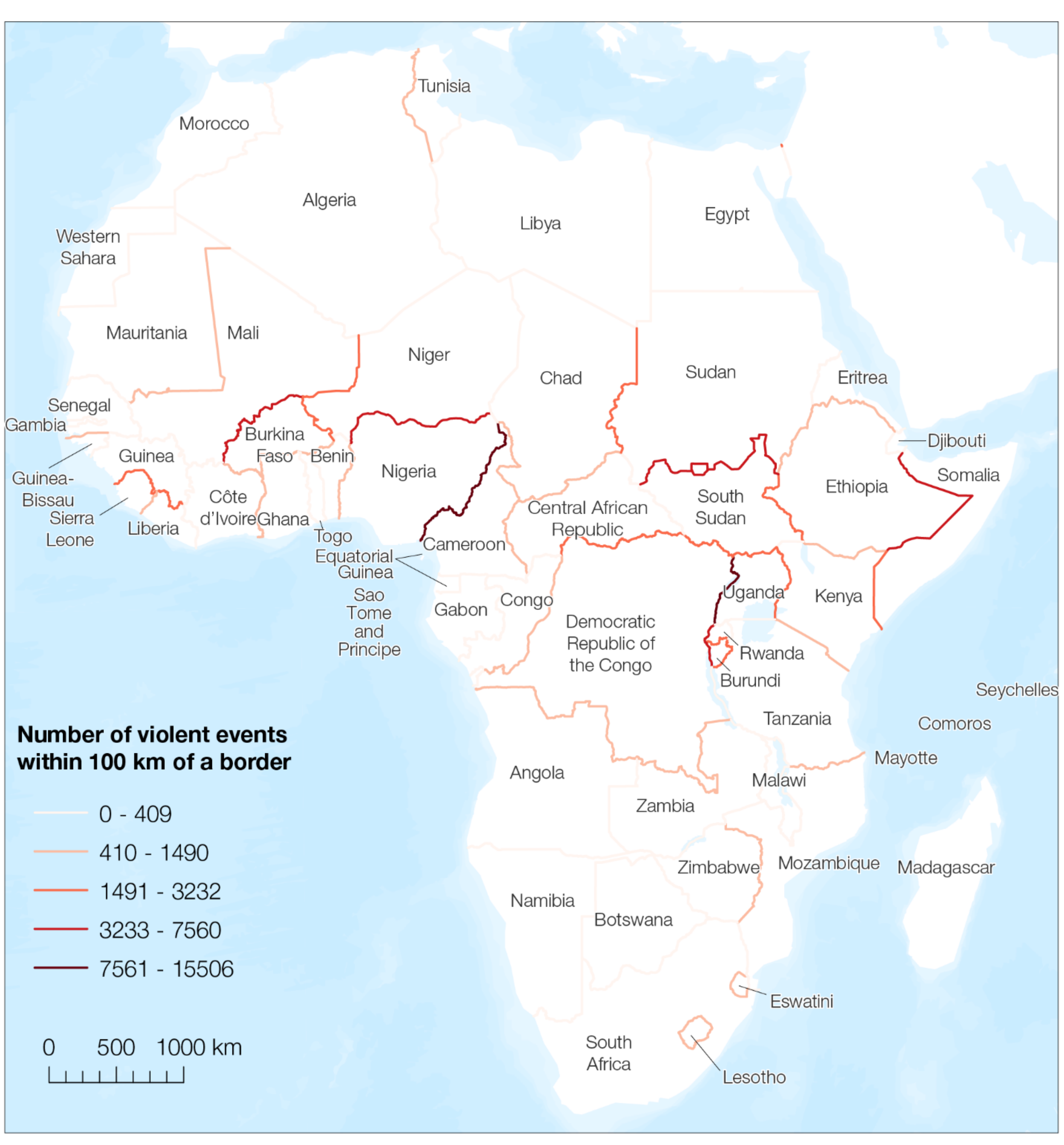

Source: ACLED (2024), calculations and cartography by the authors.

The results suggests that some border segments are much more likely to experience violence than others (Table 2). In the time period studied, the conflicts centred on the Cameroon-Nigeria border have seen the most events (more than 15,000). The north of this borderlands has been characterized by the insurgency of Boko Haram and its splinter group the Islamic State West Africa Province (ISWAP), and the counterinsurgency efforts of MNJTF based around Lake Chad. More recently, the Cameroonian side of the border has been affected by a separatist insurgency in the Anglophone region of the country.

Table 2. Most violent African borders according to conflict events, 1997-2024

| **Border** | **Within 20 km** | **Within 40 km** | **Within 60 km** | **Within 80 km** | **Within 100 km** |
|---|---|---|---|---|---|
| Cameroon - Nigeria | 4433 | 6522 | 8619 | 12748 | 15506 |
| DRC - Uganda | 2813 | 5900 | 10017 | 10803 | 11221 |
| Ethiopia - Somalia | 1066 | 2948 | 5438 | 6527 | 7560 |
| Burundi - DRC | 4691 | 5799 | 6585 | 6747 | 6827 |
| Burkina Faso - Mali | 1122 | 2520 | 3741 | 5187 | 6145 |
| DRC - Rwanda | 1968 | 3270 | 4657 | 5129 | 5291 |
| Niger - Nigeria | 1712 | 2523 | 3011 | 3567 | 4529 |
| South Sudan - Uganda | 464 | 911 | 1717 | 2921 | 3232 |
| Kenya - Somalia | 1252 | 1696 | 2053 | 2673 | 3067 |
| Sudan - South Sudan | 304 | 821 | 1914 | 2654 | 2929 |

Source: ACLED (2024), calculations by the authors.

The DRC-Uganda border is the second most affected border in terms of events (more than 11,000). This region has endured an unbroken series of conflicts since the outbreak of the Second Congo War in 1998, and the frequency of violence has intensified since 2018. Most of these conflict events have taken place in the DRC's eastern borderlands, and the violence has involved the militaries of neighbouring states as well as militias from around the region. The border between Burundi and the DRC is among the most violent segments on the continent, with nearly 7000 events. Further east, levels of political violence in Ethiopia have been high since the beginning of its Tigray War in 2020, but that conflict has mainly taken place in the far north of the country, near Ethiopia's border with Eritrea.

In West Africa, the border between Burkina Faso and Mali experienced more than 6000 violent events. These numbers reflect the expansion of the Malian conflict, which began in 2012 when Tuareg rebels took up arms to create a separatist state called Azawad. Jihadist groups took advantage of the war to launch their own campaigns in an internationalized conflict that has come to involve two French military interventions and three military coups d'etat. In the late 2010s, these jihadist groups began to take their operations southwards to neighbouring Burkina Faso, which has begun to see more conflict events than Mali (OECD, 2025).

**Political violence tends to expand rather than relocate**

Conflicts tend to spread by expansion rather than by relocation on the African continent. To visualize this, we represented the temporal evolution of the border segments that were affected by at least 4000 violent events within 100 km since 1997 (Figure 6).

Figure 6. Violent events within 100 km of African borders, by year and border segment

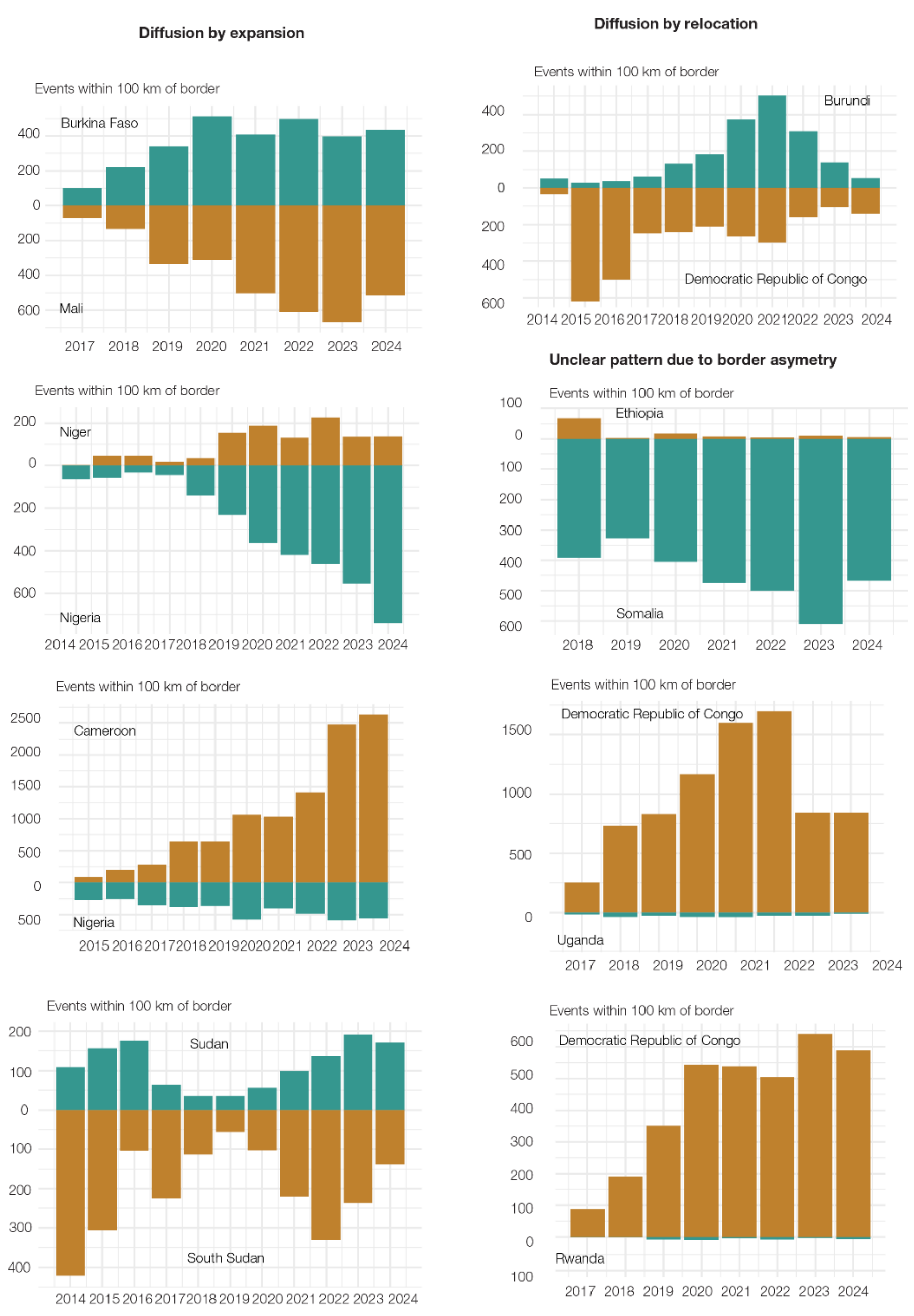

Sources: ACLED (2024), calculations by the authors. Note: An increase in the number of conflict events on both sides of the border over time indicates **expansion,** where violence spreads into new areas without diminishing in the original location; an increase in one side with a decrease on

the other indicates **relocation** of conflict from one side of the border to the other. We categorize border segments as exhibiting an "unclear pattern due to border asymmetry" when it is difficult to interpret whether conflict is expanding or relocating because the two sides of the border have experienced drastically different counts of violent events.

- In West Africa, violence between Burkina Faso and Mali, Niger and Nigeria, and Cameroon and Nigeria suggest that when conflict increases on one side of the border, they do not cease to occur at the initial place (Figure 6). In other words, conflicts in these regions tend to expand to previously unaffected areas rather than relocate completely, as evidenced by a continuous growth of violent events in both the country of origin of armed groups and their new destination. Likewise, the respective levels of violence on either side of the border between Sudan and South Sudan seems to move in tandem with each other, rather than alternating from one side of the border to the other.

- The border between Burundi and the DRC suggests a different pattern, in which a strong increase on the Burundian side of the border is met with a decrease on the Congolese side from 2015 to 2021. This would suggest a case of diffusion by relocation. In recent years, violence has decreased in the borderlands of both countries, after the government of Burundi deployed troops in Eastern Congo to combat M23 and Burundian rebels.

- Violence between the DRC, Uganda, and Rwanda is more difficult to analyse since the conflict has been so overwhelmingly concentrated on the DRC side of the border, leaving little relative variation on the other side to determine whether the pattern reflects expansion or relocation. A similar trend is observed between Ethiopia and Somalia. The vast majority of conflict events that have occurred in these borderlands took place on the Somalian side of the border due to the intense civil war that has occurred and recurred in this country since the early 1990s.

A more robust approach to conflict diffusion is to calculate correlation tests on all pairs of borders. If the correlation coefficient is strongly negative, it would indicate relocation, and if it is strongly positive, it would indicate expansion. The analysis of the top 27 most violent borders by number of events within 100 km from 1997-2024 reveals that there are 14 clear cases of diffusion by expansion and only 2 cases of diffusion by relocation: Burundi-DRC from 2014 to 2021 and the Eritrea-Ethiopia War from 1998 to 2000, which is one of the very few interstate conflicts included in the analysis. The rest of the border regions have low coefficients, suggesting that there is hardly any relation between the two countries, or that conflict really only takes place on one side of the border. This method is sensitive to how the time period of interest is defined, however, since the geography of a cross-border conflict can change over time.

This approach captures the transnational dimension of a conflict region, not the actual spatial patterns of state or non-state actors. In other words, while the approach can distinguish between conflicts that tend to relocate or expand, it cannot follow the movements of belligerents across borders. This is particularly limiting in border regions where several violent groups are fighting against the government. Only a disaggregated approach conducted at the scale of individual groups could potentially explain *why* violence is expanding or relocating.

**Perspectives**

The tendency of African conflict to expand emphasises the need to address borderlands on both sides of the border as an interconnected whole in conflict studies. Both sides of a border may experience violence simultaneously for several reasons: rebels and extremists may share the same grievances vis-à-vis the central government or may want to take advantage of the political insecurity created in the neighbouring country to launch their own movement.

Thus far, the spatial and temporal dynamics that sustain these armed conflicts in borderlands remain understudied despite their importance to transnational security concerns on the African continent. This is due to several challenges.

The first challenge lies in dealing with inconsistent data, fragmented by country of origin. This study points to a possible solution to the data disconnect. Research on political violence has benefited tremendously from the proliferation of spatially disaggregated event datasets such as ACLED. In addition to allowing for a more granular study of conflict at sub-national levels, these datasets encourage a transnational approach at larger scales. When data is spatially disaggregated and gathered according to continent-wide standards, approaches to cross-border phenomena can become more methodologically consistent. They can also emphasise the transnational, interrelated context on which border studies should centre, allowing for border studies to inform more internationally coordinated policy responses.

A second challenge lies in incorporating the interactions between borders and other spatial features. Certain parts of borderlands are given importance by the transportation network that traverses and is circumscribed by them. Likewise, the economies of towns and cities that straddle borders or are near to border crossings is heavily shaped by cross-border trade. Taken together, roads, cities and borders shape the spatial dynamics of conflict observed in Africa: violence is more intense near them and decreases sharply as one moves away from populated centres, transport corridors, and international boundaries (OECD, 2022; 2023; 2025).Linking patterns of political violence to these spatial features can also help promote an integrated understanding of how political violence affects development issues such as regional integration.

The robust relationships between these spatial features and armed conflicts suggest that there are regularities in the way violence is spatially and temporally distributed in Africa (Walther et al., 2025). Future studies should build on these regularities to develop a more comprehensive understanding of the trajectory of armed conflicts and model how violence diffuses across the continent over space and through time.

Finally, more work needs to be done to explain the interface between the marginality of borders and the centrality of the hegemonic nation-state system. The reasons for crossing a border might have as much to do with the presence of the border as it does with the state that creates the border. While borderlands are often marginalized, they are also sites where state power is most reified.

Future research should be dedicated to understanding how the expansion of armed conflicts in African borderlands contributes to reshape the social contract established between the central government and borderlanders. The case of the Sahel suggests that state initiatives have mainly focused on strengthening security and restraining mobility, rather than an integrated approach that encourages trade, rehabilitates infrastructure, or facilitates social exchanges alongside increasing security capacity (OECD, 2025). This militarization of African borderlands, which has largely taken precedence over developmental support for border communities, deserves further scrutiny as it undermines long-term regional stability